\documentclass[iop]{emulateapj}
\usepackage[english]{babel}
\usepackage[utf8x]{inputenc}
\usepackage{amsmath}
\usepackage{graphicx}
\usepackage[colorinlistoftodos]{todonotes}

\newcommand{\taud}{\tau_{\rm d}}

\newcommand{\tsp}{\sigma}

\newcommand{\DMunits}{\mbox{pc cm}^{-3}}
\newcommand{\tauGHz}{\tau_{\rm 1GHz}}

\newcommand{\thetaobs}{\theta_{\rm o}}

\newcommand\pasa{\ref@jnl{PASA}}%
\begin{document}
\title{Pulse Broadening Measurements from the Galactic Center Pulsar J1745--2900}

\author{L.~G.~Spitler,\altaffilmark{1} K.~J.~Lee,\altaffilmark{1} R.~P.~Eatough,\altaffilmark{1} M.~Kramer,\altaffilmark{1,2} R.~Karuppusamy,\altaffilmark{1} C.~G.~Bassa,\altaffilmark{2} 
I.~Cognard,\altaffilmark{3} G.~Desvignes,\altaffilmark{1} 
A.~G.~Lyne,\altaffilmark{2} B.~W.~Stappers,\altaffilmark{2} 
G.~C.~Bower,\altaffilmark{4} J.~M.~Cordes,\altaffilmark{5} D.~J.~Champion,\altaffilmark{1} \& H.~Falcke,\altaffilmark{1,6,7}}
\affil{$^1$Max-Planck-Institut f\"ur Radioastronomie, Bonn, 53121, Germany}
\affil{$^2$Jodrell Bank Centre for Astrophysics, School of Physics and Astronomy, The University of Manchester, M13, 9PL Manchester, UK}
\affil{$^3$Laboratoire de Physique et Chimie de l'Environnement et de l'Espace LPC2E CNRS-Université d'Orléans, F-45071 Orléans Cedex 02, and Station de radioastronomie de Nan\c{c}ay, Observatoire de Paris, CNRS/INSU, F-18330 Nan\c{c}ay, France}
\affil{$^4$UC Berkeley Astronomy Dept, B-20 Hearst Field Annex, Berkeley, CA 94720-3411}
\affil{$^5$Department of Astronomy and Space Sciences, Cornell University, Ithaca, NY 14853}
\affil{$^6$Department of Astrophysics, Institute for Mathematics, Astrophysics and Particle Physics, Radboud University, PO Box 9010, 6500 GL Nijmegen, The Netherlands}
\affil{$^7$ASTRON, P.O. Box 2, 7990 AA Dwingeloo, The Netherlands}

%COMMENTS
%REA ET AL - Jim: Not convincing on the Chandra spectra; Me: Look over (newer version)
%Scattering section - Jim: PBF is a time / freq-averaged results; Me: not sure what to say
%Results - Jim: avoid color in figures

\begin{abstract}
We present temporal scattering measurements of single pulses and average profiles of PSR J1745--2900, a magnetar recently discovered only 3 arcsec away from Sagittarius A* (Sgr A*), from 1.2 - 18.95 GHz using the Effelsberg 100-m Radio Telescope, the Nan\c{c}ay Decimetric Radio Telescope, and the Jodrell Bank Lovell Telescope. Single pulse analysis shows that the integrated pulse profile above 2 GHz is dominated by pulse jitter, while below 2 GHz the pulse profile shape is dominated by scattering. The high dispersion measure and rotation measure of the magnetar suggest that it is close to Sgr A* (within $\sim$0.1 pc). This is the first object in the GC with both pulse broadening and angular broadening measurements. We measure a pulse broadening spectral index of $\alpha = -3.8 \pm 0.2$ and a pulse broadening time scale at 1 GHz of $\tauGHz = 1.3 \pm 0.2$ s, which is several orders of magnitude lower than the scattering predicted by the NE2001 model \citep{cl02}. If this scattering timescale is representative of the GC as a whole, then previous surveys should have detected many pulsars. The lack of detections implies either our understanding of scattering in the GC is incomplete or there are fewer pulsars in the GC than previously predicted. Given that magnetars are a rare class of radio pulsar, we believe that there many canonical and millisecond pulsars in the GC, and not surprisingly, scattering regions in the GC have complex spatial structures. 
\end{abstract}

\section{Introduction}
The recent discovery of radio pulsations from a magnetar at a projected distance of $0.12$ pc from Sagittarius A* (Sgr A*) provides an unparalleled tool for probing the ionized interstellar medium (ISM) toward the Galactic Center (GC) \citep{efk+13}. 
SGR J1745--29 was first identified at X-ray wavelengths by the {\em Swift} observatory during regular monitoring of Sgr A* \citep{kbk+13}. Targeted follow-up observations by the {\em NuSTAR} observatory revealed pulsed emission with a period of 3.76 s and a high period derivative\citep{mgz+13}. The inferred magnetic field of $\sim 10^{14}$ G, X-ray spectral properties, and sudden increase in flux suggests that the object is a transient magnetar \citep{mgz+13}. 

Radio follow-up observations of SGR J1745-29 (hereafter PSR J1745--2900) have confirmed the high spin down rate, $(6.82 \pm 0.03) \times 10^{-12}$, and yielded a dispersion measure (DM) of DM = 1778 $\pm$ 3 $\DMunits$ \citep{efk+13}, the highest DM of any known radio emitting neutron stars. The NE2001 model for the distribution of Galactic electrons \citep{cl02} estimates a DM distance for the magnetar that is consistent with the distance of Sgr A*. \citet{efk+13} measure a high rotation measure (RM) of ${\rm RM} = -66960 \pm 50\;{\rm rad\;m^{-2}}$ \citep[see also][]{sj13}. This RM is an order of magnitude larger than all other RMs measured within a few tens of parsecs of Sgr A* \citep[see e.g.][and references therein]{lbn11}, and \citet{efk+13} postulate it is caused by the
hot gas component from which Sgr A* accretes (starting at the Bondi radius of $\sim$0.12 pc).
Furthermore, \citet{mgz+13} find that the column density inferred from the X-ray spectrum measured by \emph{NuSTAR} is consistent with the magnetar being at the distance of the GC, and \citet{rep+13} compare the column densities measured from X-ray spectra of Sgr A* and PSR J1745--2900 from the \emph{Chandra X-ray Observatory} and infer an upper limit to their physical separation of $\lesssim$ 2 pc. 

A surprising result of the radio measurements has been the small pulse broadening due to interstellar scattering of radio waves. The NE2001 model predicts a large pulse broadening timescale of $\sim$ 2300 s for a source in the GC. Previously the lack of pulsars observed in the GC, despite strong evidence for their existence in this region \citep[see e.g.][]{wcc+12}, was explained by extreme scattering of radio waves caused by inhomogeneities in the ionized component of the interstellar medium within $\sim$150 pc of Sgr A* \citep{lc98b}. Scattering causes temporal broadening of pulses which renders pulsar periodicity searches at typical observing frequencies ($\sim 1$ GHz) ineffective.  

In this letter, we present detailed multi-frequency measurements of the temporal broadening from PSR J1745--2900 and discuss their implications. In Section~\ref{scattering} we give a brief description of the observational phenomena of pulse and angular broadening. Our observations and data analysis are given in Section~\ref{obs}. The results are presented in Section~\ref{results}. We discuss our results and summarize in Sections~\ref{discussion} and \ref{conclusions} respectively.

\section{Interstellar Scattering}
\label{scattering}
A pulse propagating through the non-uniform, ionized ISM will be scattered, leading to the multi-path propagation effects of angular broadening and pulse broadening. A point source is broadened to a typical observed angular size $\thetaobs$, and an impulse-shaped pulse is broadened to a characteristic time scale $\taud$. These two quantities, $\thetaobs$ and $\taud$ are geometrically related and depend on the properties of the scattering screen \citep[see e.g.][]{w72,r90}.

Mathematically, temporal scattering is described by the pulse broadening function (PBF). The observed pulse shape is the convolution of the intrinsic pulse shape and the PBF. The most commonly assumed geometry is a thin screen with infinite transverse extent dominated by the smallest spatial scale, which gives a one-sided exponential PBF \citep{c70}:
\begin{equation}
	{\rm PBF_e}(t) = \Theta(t) e^{-t/\taud} 
\end{equation}
where $\Theta(t)$ is the unit step function, i.e.\ $\Theta(t\ge0)=1$, otherwise 
$\Theta(t)=0$.  If the scattering medium is instead a thick screen, the PBF has a slower rise time than ${\rm PBF}_{\rm e}$ \citep[e.g.][]{w72, bcc+04}. Kolmogorov media have PFBs that decay more slowly than an exponential \citep{lr99, cr98}. Other possible geometries include scattering screens with limited transverse extent or filaments \citep{cl01}. 

Pulse and angular broadening is also highly frequency-dependent with a typical spectral index $\alpha$ proportional to $\sim \nu^{-4}$ and $\sim \nu^{-2}$ respectively. \citet{bcc+04} measured a mean spectral index of $\alpha = -3.9 \pm 0.2$ for a large ensemble of pulsars with low to moderate DMs. \citet{lkm+01} measured the pulse broadening spectral index of nine pulsars with large DMs ($\sim 400-1000\; \DMunits$) and determined a shallower frequency scaling of $\alpha = -3.44 \pm 0.13$.

\section{Observations and Data Analysis}
\label{obs}
The magnetar was observed with the Effelsberg radio telescope at observing frequencies ranging from 1.35 to 18.95 GHz, the Nan\c{c}ay radio telescope from 1.5 - 3.2 GHz, and the Jodrell Bank Lovell telescope at 1.5 GHz. A full list of observing frequencies, bandwidths, and observing dates is given in Table~\ref{tab:dat}.

\subsection{Effelsberg}
The 18.95 GHz data were taken with the P13mm receiver and XFFTS digital spectrometer \citep{khk+12}. The XFFTS produces spectral intensity data over a bandwidth of 2 GHz with 256 frequency channels and a time resolution of 128 $\mu$s. The 14.6 GHz, 8.35 GHz, and 4.6 GHz data were taken with the Effelsberg S20mm, S36mm, and S60mm receivers respectively. At 1.4 GHz both the Ultra-broadband receiver (UBB) and the central pixel of the 21-cm multi-beam receiver (7B) were used. At the four lower frequencies baseband data were recorded with the PSRsterix coherent dedispersion system. 

The Effelsberg data at 18.95 GHz were dedispersed at DM = 1778 $\DMunits$ using the PRESTO\footnote{\tt{http://www.cv.nrao.edu/~sransom/presto/}} pulsar processing software suite. 
The dedispersed time series were convolved with a set of boxcar filters of increasing width to increase the sensitivity to broader pulses. Pulses were identified by applying a minimum signal-to-noise ratio threshold of S/N$>$5 to the convolved dedispersed time series and binned by pulse phase. A total of 21 pulses with S/N$>$5 were seen in the on-pulse phase window at boxcar widths ranging from 128 $\mu$s -- 768 $\mu$s. Because the pulses were too weak to fit for scattering individually, a ``dejittered" average profile was created by co-adding short segments of the dedispersed time series centered on each detected pulse. 

At all other observing frequencies at Effelsberg, the processing went as follows. The baseband data were coherently dedispersed, and for each rotation of the pulsar, spectral intensity data were generated. An average pulse profile was generated by integrating over the entire observation. At 4.85 and 8.35 GHz a single frequency subband was used, while at 1.4 GHz data two and four frequency subbands were made for the 7B and UBB data respectively. Data reduction and RFI excision were performed using the standard psrchive package \citep{hsm04}.

Single pulse time series were generated at 4.85, 8.35, and 14.6 GHz for each pulsar rotation by first averaging over frequency. We perform further processing to avoid fitting pulses caused by short-duration radio frequency interference (RFI) spikes. First we identify the phase window during which the magnetar was ``on", and assume that the phase bin with the maximum flux is a single pulse. A short segment of the time series centered on the pulse is extracted and used for the single pulse profile fitting described in Section~\ref{spfit}. 

\subsection{Nan\c{c}ay}
Observations at the Nan\c{c}ay Radio Telescope were taken using the
NUPPI instrument at three different central frequencies: 1.48 GHz with the Low Frequency receiver and 2.54 GHz and 3.18 GHz with the High Frequency receiver. At all three frequencies, a bandwidth
of 512 MHz was split into 1024 channels, coherently dedispersed at a DM
of 1830 pc cm$^{-3}$ and subsequently folded at the initially measured spin
period. Because the data were folded in realtime, no single pulses measurements were available.

\subsection{Jodrell Bank}
At Jodrell Bank, observations were performed using the Lovell Telescope
at a central frequency of 1532 MHz, using a 350 MHz wide band divided
into 0.5 MHz channels. A dual-polarization cryogenic receiver was used
and orthogonal circular polarization were recorded using a coherent dedispersion system which processed the raw voltages in real time. No single pulses archives were generated for these data, because they were folded in real time. 

\begin{table*}
\centering
	\caption{Observational parameters and best fitted parameters
	\label{tab:dat}}\begin{tabular}{c c c c c c l }
		\hline \hline
Frequency & Bandwidth & Epoch & Length &$\tau$ & $\sigma$ & Comments \\
(GHz)     & (MHz)     &       & (hr)   & (ms)  &   (ms)   &         \\ \hline
8.36 & 500&    2013-05-06&	1.1 &  4.8$\pm$3.5 &  32$\pm$2    &   EFF \\
4.86 & 500&    2013-06-14&	1.2 &  5.8$\pm$5.3 &	37$\pm$2     &   EFF\\
3.18 & 512&    2013-06-18&	0.4 & 18$\pm$6      &    47$\pm$3   &    NCY\\
2.56 & 512&    2013-06-17&  0.5 & 47$\pm$6      &   55$\pm$4    &   NCY \\
2.56 & 512&    2013-06-19&	0.5 & 25.3$\pm$5    &     50$\pm$3  &     NCY \\
1.63 & 128&    2013-07-19&  0.5 & 362$\pm$35    &    83$\pm$15  &    NCY \\
1.63 & 192&    2013-06-24&	6.4 &  229$\pm$13    &    39$\pm$4   &    JB \\
1.55 & 128&    2013-07-19&  0.5 & 214$\pm$15    &    81$\pm$11  &   NCY\\
1.46 & 192&    2013-06-24&	6.4 &  292$\pm$28    &    47$\pm$10  &    JB \\
1.42 & 128&    2013-07-19&     0.5 & 365$\pm$28    &    60$\pm$10  &   NCY\\
1.42  &125& 2013-07-26 &       2.1 &  263$\pm$22&     $79\pm10$     & EFF (7B) \\
1.42 & 78&     2013-07-25&	2.2 & 383$\pm$61     & 91$\pm$25    &   EFF (UBB)\\
1.34 & 78&     2013-07-25&	2.2 & 564$\pm$80     & 42$\pm$18    &  EFF  (UBB)\\
1.30 & 128&     2013-07-26&	2.1 & 605$\pm$43   &     87$\pm$14&      EFF (7B)\\
1.29 & 128&    2013-07-19&	0.5 &  488$\pm$51    &    47$\pm$14  &   NCY\\
1.27 & 78&     2013-07-25&	2.2 & 531$\pm$95     &  164$\pm$40   &   EFF (UBB)\\
1.19 & 78&     2013-07-25&	2.2 & 1423$\pm$320   &  54$\pm$25    &    EFF (UBB)\\
\hline                          
4.86 & 500&    2013-06-14&  1.2 &   3.3$\pm$0.6 &   6$\pm$2     &   EFF$\rm ^s$ \\
8.36 & 500&    2013-05-06&  1.1 &  0.3$\pm$0.4  &  1$\pm$0.8   &   EFF$\rm ^s$ \\
14.6 & 500&    2013-06-14&  1.2 &  0.25$\pm$7   & 2$\pm$6      &   EFF$\rm ^s$\\
18.95 & 2000&  2013-05-07&  2.4 &   0.2$\pm$0.07 & 0.12$\pm$0.04&   EFF$\rm ^s$\\
\hline \hline
	\end{tabular}
    
The `EFF', `NCY', and `JB' refer to Effelsberg 100-m radio telescope, Nan\c cay Decimetric Radio Telescope, and Jodrell Bank Lovell Telescope. `7B' and `UBB' denote the seven beam and the ultra-broad band receiver at Effelsberg telescope. The superscript `S' denotes single pulse data. The $\pm$ sign indicate the 1-$\sigma$ standard deviation of the parameters.
\end{table*}

\section{Results}
\label{results}

\begin{figure}
\includegraphics[scale=0.325,angle=-90]{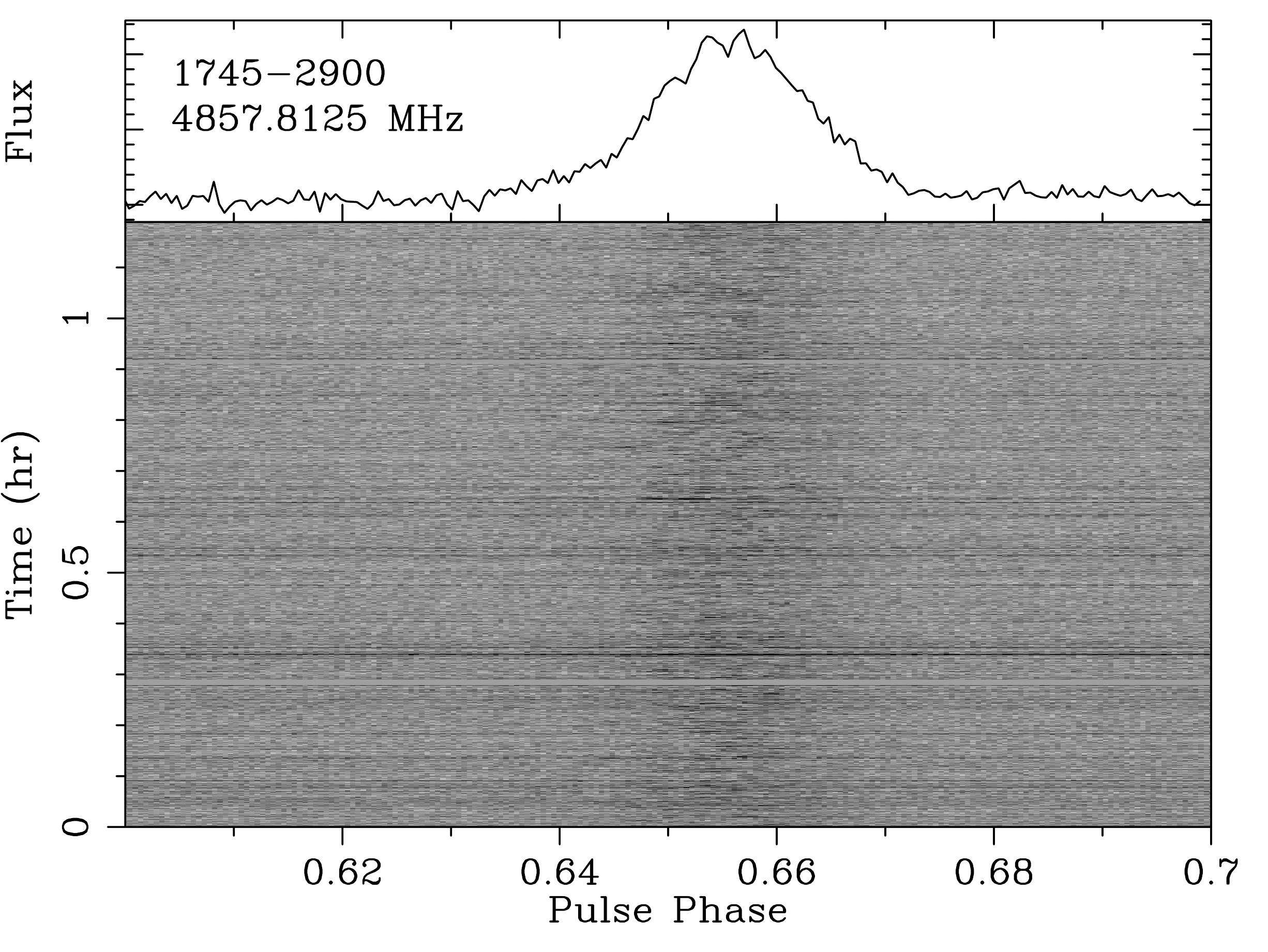}
\includegraphics[scale=0.325, angle=-90]{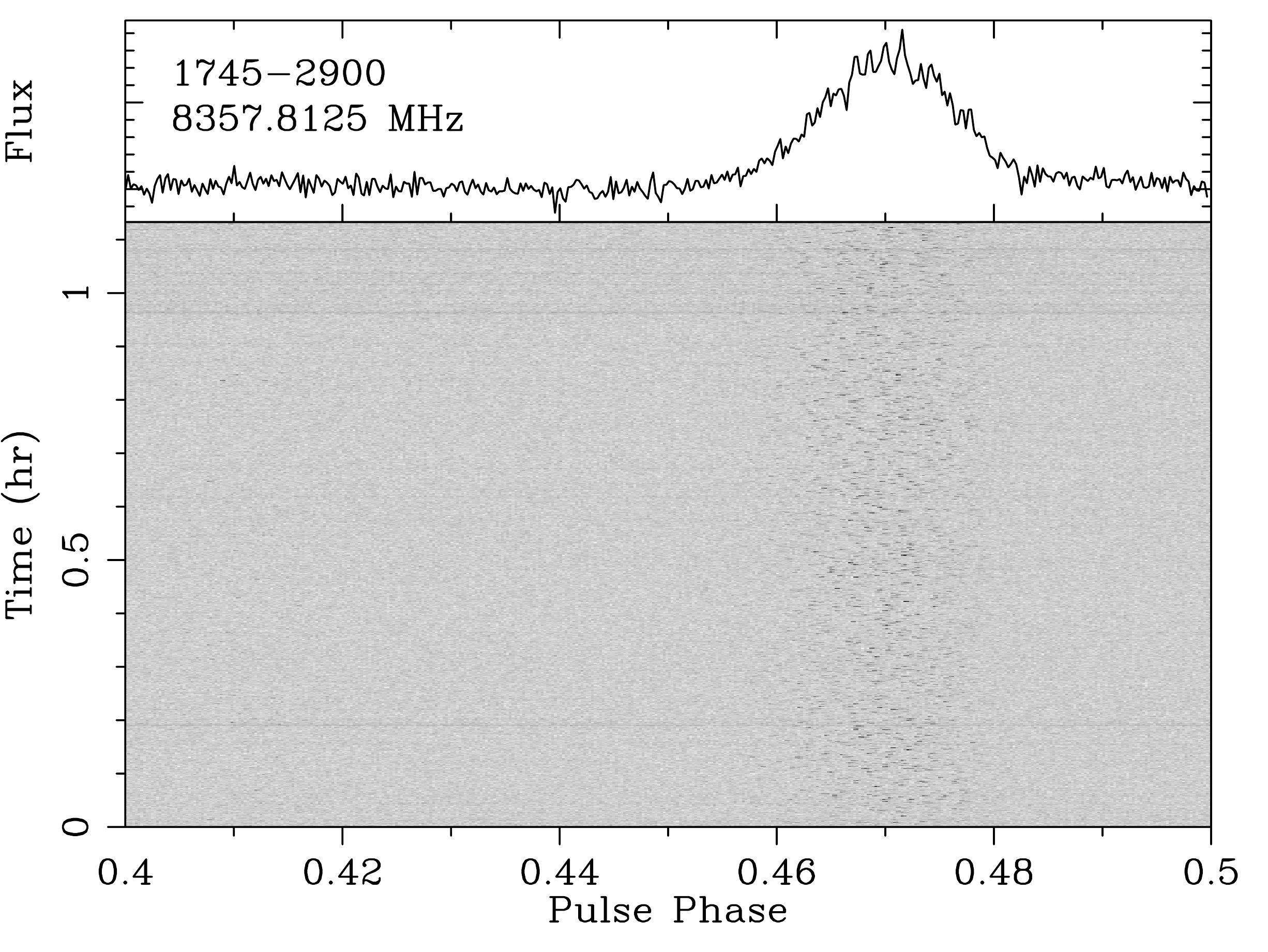}
 \caption{Flux vs.\ pulse phase plots from Effelsberg observations of PSR J1745--2900 showing pulse jitter at 4.6 GHz and 8.35 GHz. The lower panel of each figure shows flux vs. pulse phase and observation time, and the top panel shows the pulse profile averaged over the entire observation. The rotation-resolved profiles clearly show that the integrated profile is comprised of many narrower single pulses with stochastically varying phases. }
 \label{fig:f1}
\end{figure}

The single pulse analysis revealed that the emission beam of the magnetar consists of one or more narrow pulses with widths of $\sim$ 1 ms. The phases of single pulses vary stochastically over a range that is roughly an order of magnitude larger than the measured intrinsic widths of the single pulses. The average pulse profile is therefore dominated by jitter, a behavior seen in other magnetars \citep{crh+06, ksjl07, lbb+12}. Figure~\ref{fig:f1} shows integrated pulse profiles (top panels) and pulse profiles from individual rotations (bottom panels) for data taken with Effelsberg at 4.85 and 8.35 GHz. The shapes of the average profiles are also Gaussian-like, which is important for our pulse  profile fitting technique described in Section~\ref{sec:fit}. 

The pulse profiles of PSR J1745--2900 are characterized by three time scales. The shortest is the width of the single pulses. The phase jitter timescale is longer than the temporal widths of the single pulses and dominates the timescale of average profile at frequencies above $\sim$ 2 GHz. The third timescale is the pulse broadening caused by interstellar scattering and dominates the average profile at frequencies below $\sim$ 2 GHz.

\begin{figure}
\includegraphics[totalheight=4.5in]{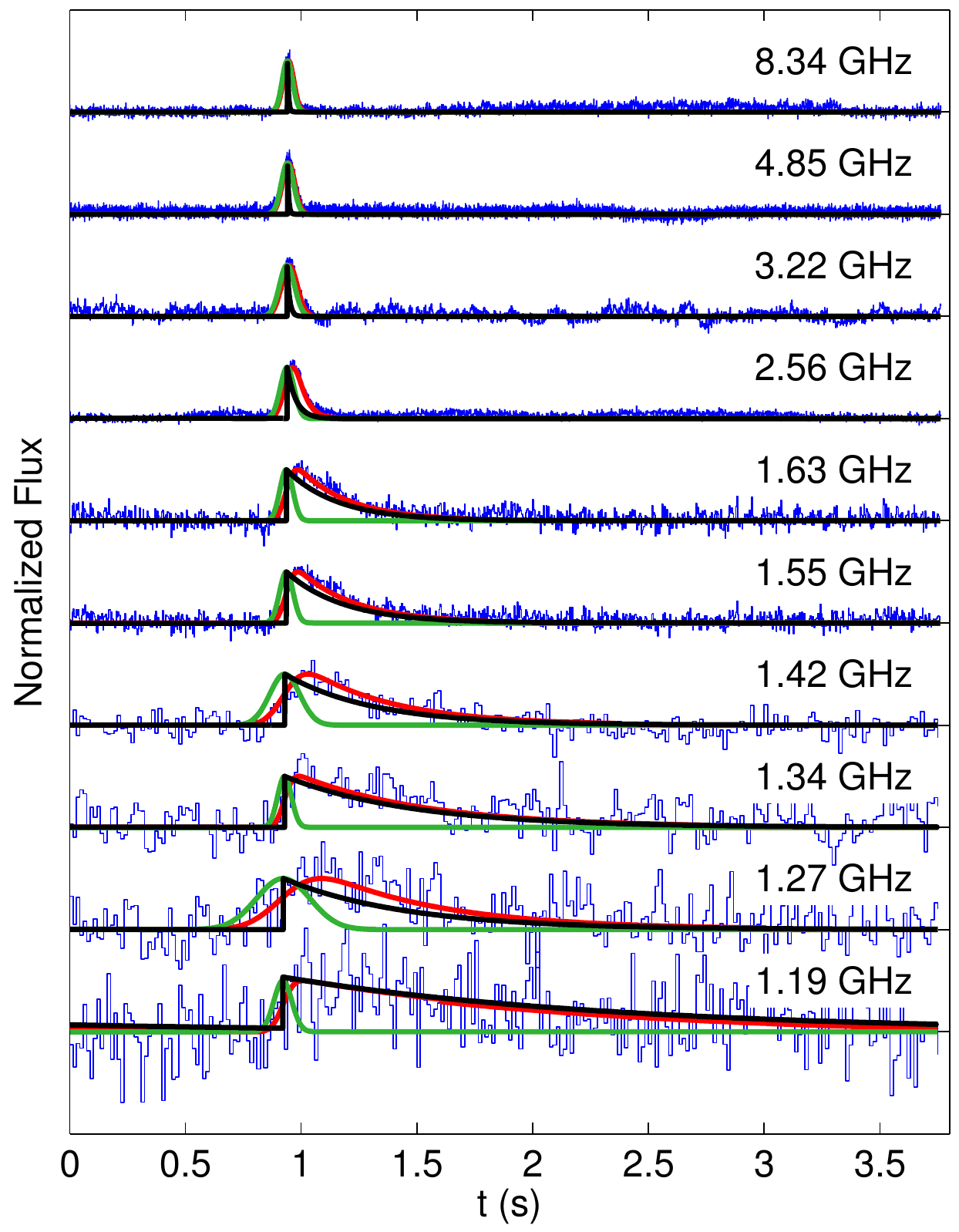}
 \caption{Multi-frequency integrated pulse profiles. The blue curves are the measured profile. The red, green and black lines are the best fitted profile $P_{\rm obs}(t)$, best-fit Gaussian profile $P_{\rm g}(t)$ and the scattering filter ${\rm PBF_{\rm e}}(t)$ respectively.  
\label{fig:prof}}
\end{figure}

\subsection{Model for Pulse Shape}
\label{spfit}
We model both single pulses and average profiles as scattered Gaussian pulses that have the same scattering time $\taud$ but different Gaussian widths $\sigma$, i.e.\ 
\begin{equation}
	P_{\rm g}(t) = A e^{-\left(\frac{t-t_0}{\tsp}\right)^2},
\end{equation}
where $A$ and $t_0$ are the amplitude and epoch of the pulse peak. 
For single pulses the parameter $\sigma$ represents the intrinsic width of the pulses, while for the average profiles it represents the jittered width.  

The scattered pulse profile $P_{\rm s}(t)$ is the intrinsic profile convolved with the PBF. That is
\begin{eqnarray}
	P_{\rm s}(t) &=&P_{\rm g}(t) * {\rm PBF_e}(t)\nonumber\\
	&=&\frac{A\sqrt{\pi}}{2}\tsp e^{\displaystyle \frac{\tsp^2-4 t \taud}{4\taud^2}} 
	\left[1+\textrm{erf}\left( \frac{2 (t-t_0)\taud -\tsp^2 }{2 \taud 
	\tsp}\right)\right],
\end{eqnarray}
 where the asterisk denotes convolution.
The instrumental response further modifies the pulse profile. 
The observed pulse profile is thus
\begin{equation}
	P_{\rm obs}(t) =P_{\rm s}(t) * S(t)  +b\,,
	\label{eq:pobs}
\end{equation}
where $b$ is the profile baseline level, and the integration-sampling function 
$S$ is
\begin{equation}
	S(t)=\left\{\begin{array}{c c}
		1 & -\tau_{\rm s}\le t\le0, \\
		0 & \textrm{otherwise},
	\end{array}
	\right.
\end{equation}
where $\tau_{\rm s}$ is the sampling time for the pulse profile.

$P_{\rm obs}(t)$ should also include the residual dispersion measure smearing across a frequency channel $d(t)$ \citep[e.g.][]{cm03}.  For the coherently dispersed data, the factor $d(t) = 0$.  At 18.95 GHz the dedispersion smearing across a frequency channel is 16.9 $\mu$s, which is much less than the sampling time of 128 $\mu$s and can be ignored. 

\subsection{Pulse profile fitting}
\label{sec:fit}
We fit the observed single pulse profiles and the integrated pulse profiles with the models described above.  
The pulse profile is denoted with the parameters $\{t_i, p_i,\sigma_i\}$, where $t_i$, $p_i$, and $\sigma_i$ are the time, 
pulse flux and pulse flux error of the $i$-th data point. The index $i$ goes from 1 to 
$N_{\rm b}$, the number of bins in the pulse profile. The  
reduced $\chi^2$ for the five parameter fit is
\begin{equation}
	\chi^2=\frac{1}{N_{\rm b}-6} \sum_{i=1}^{N_{\rm b}} \left[\frac{P_{\rm 
	obs}(t_i; A, t_0, \tsp, b, \tau_{\rm d})-p_i}{\epsilon_i}\right]^2.
\end{equation}
The error $\epsilon_i$ is estimated via the off-pulse RMS level. 
The least squares solution for the parameters was found by minimizing the $\chi^2$ using a down-hill simplex method.
We seed the initial fitting parameters randomly and repeat for 100 unique trials in order to find the global minima.  
Our error convention follows \cite{BW07}, the 1-$\sigma$ error is defined as the square root of diagonal terms of the covariance matrix for the fitting residuals. 

We fit the single pulse profiles at 4.85, 8.35, 14.6, and 18.95 GHz, and the integrated pulse profile at 1.4, 2.5, 3.1, 4.85, and 8.35 GHz.  The best-fit parameters of all profile fits are given in Table~\ref{tab:dat}. We did not fit for the scattering timescale in the 14.6 GHz and 18.95 GHz averaged profile, since it is nearly 1000 times smaller than the pulsar jitter time scale at 
these frequencies. No systematical structure is found in the fitting residuals, which indicates that the Gaussian modeling for the intrinsic profile is sufficient. Examples of observed pulse profiles from frequencies of 1.19~GHz to 8.34~GHz (blue lines) and their best fit model profiles are given in Figure~\ref{fig:prof} with the red lines showing the convolved model profile $P_{\rm obs}(t)$, and green the best-fit Gaussian profile $P_{\rm g}$, and black the ${\rm PBF_e}$. 

\begin{figure*}
\centering
 \includegraphics[totalheight=4in]{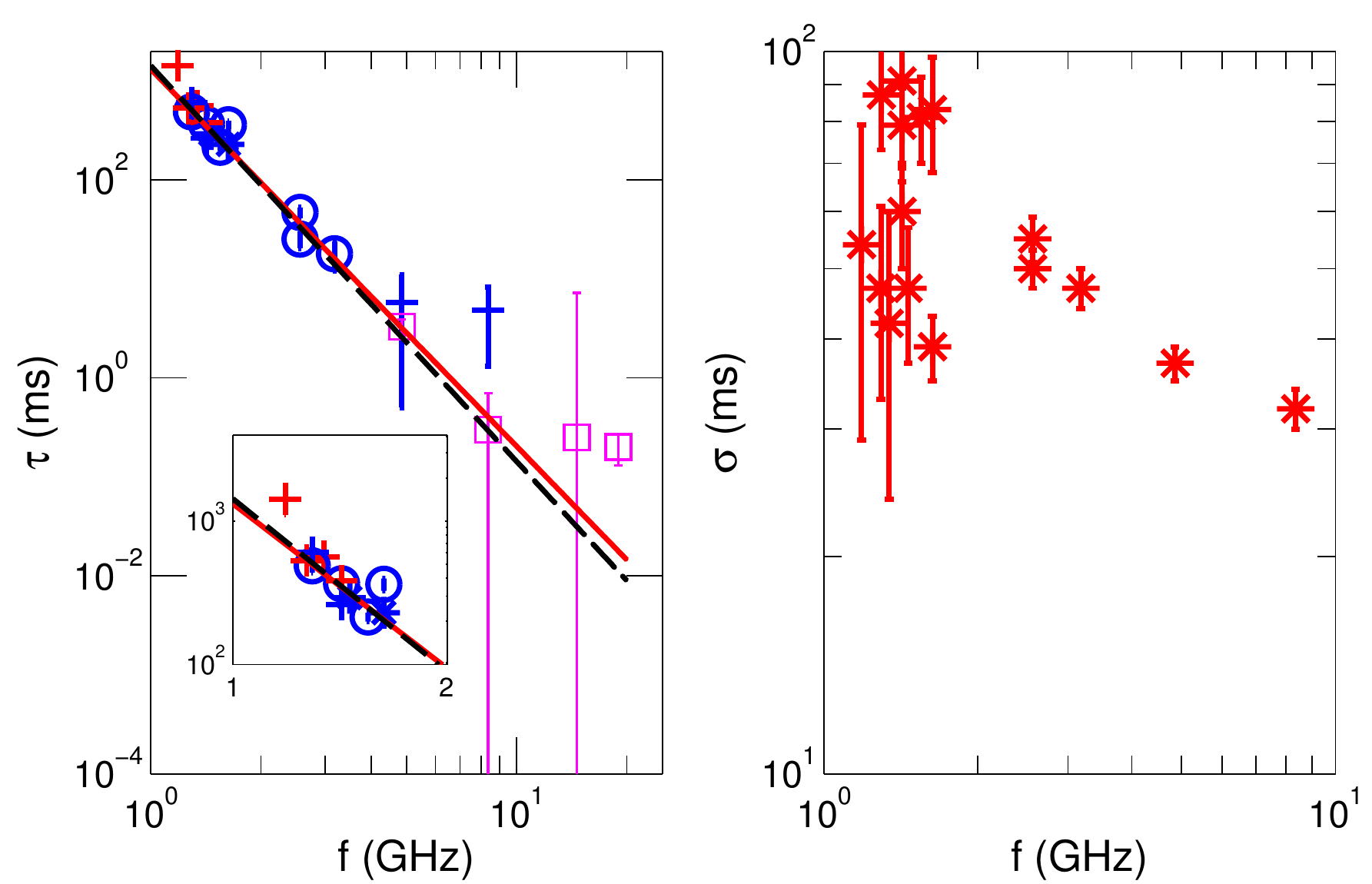}
 \caption{Measurements of the scattering broadening timescale ($\taud$) and intrinsic pulse width ($\sigma$) of PSR J1745--2900 from 1.2 to 18.95 GHz together with the 1-$\sigma$ errorbar. Left panel: The measured pulse scattering time scales $\tau_{\rm d}$ as a function of observing frequency $\nu$. The $+, \circ$, and $*$ denotes the data from Effelsberg, Nan\c{c}ay, and Jodrell Bank respectively. The purple squares are from Effelsberg single pulse data. The inset in the bottom left shows a zoomed region for 1 to 2 GHz. The red solid line is a simultaneous fit for the pulse broadening timescale and spectral index, which yields a scattering timescale at 1 GHz of $\tauGHz$ =  1.3$\pm$0.2 s and power-law index of $-3.8\pm 0.2$. The black dashed line is a fit fixing the power index to $-4$, which gives $\tauGHz = 1.4\pm0.1$ s.
Right panel: The best-fit Gaussian widths ($\sigma$) of the averaged pulse profiles as a function of frequency. The measurements between 2 and 20 GHz may suggest that the jitter-dominated time scale may vary in frequency, but the large scatter in the 1 to 2 GHz intrinsic widths makes it difficult to draw firm conclusions.
}
\label{fig:fit}
\end{figure*}

Figure~\ref{fig:fit} shows the scattering time scale ($\taud$) versus frequency derived from both the integrated profiles and single pulses in addition to the intrinsic pulse width ($\tsp$) versus frequency for the integrated profiles only. See the figure caption for a description of the symbols. The lowest frequency measurement from Effelsberg shows a greater than 1-$\sigma$ deviation from the scattering function. As measurements at the same frequency from Nan\c{c}ay do not show this deviation, we suspect it is caused by a baseline effect. 
The apparent deviation of $\taud$ at 18.95 GHz likely reflects the finite sampling time of the data. The dejittered average profile is just two sampling intervals wide; the fitted parameters are of the same order as the 128~$\mu$s sampling time. 

A least squares fit of all of the scattering timescales given in Table~\ref{tab:dat} versus frequency yielded a scattering spectral index of $\alpha$ = $-3.8\pm 0.2$, which is consistent with the value of $-3.9\pm0.2$ determined by \citet{bcc+04} and $-3.44\pm0.13$ from \citet{lkm+01}. The projected scattering timescale at 1 GHz is $\tauGHz$ = 1.3$\pm$0.2 s.

\section{Discussion}
\label{discussion}
PSR J1745--2900 is the only astronomical object in the GC for which both angular scatter broadening and pulse broadening have been measured, which can constrain the location of the scattering medium \citep{cl97} . The angular size of PSR J1745--2900 measured by Bower \emph{et al., submitted} of $16.0^{+1.1}_{-1.1} \times 9.7^{+1.6}_{-1.9}$ mas and our measurement of $\tauGHz$ = 1.3$\pm$0.2 s place the screen at a distance of $\approx$ 6 kpc from the GC (see Bower \emph{et al., submitted} for details). While it is plausible that an HII region in a spiral arm along the line of sight could cause strong scattering, it also implies the scattering in the GC is much lower than previously thought. If this is true, then many more pulsars in the GC should have been detected in previous search attempts \citep{kkl+00, jkl+06, dcl09, mkf+10, ekk+13}.

One possible resolution to this apparent contradiction is that the thin screen scattering model is invalid for sources in the GC. \citet{lc98b} present a more realistic two-component model with a central spheroid of hot gas and a scattering screen located $\sim$ 150 pc from Sgr A*. The physical origin of the scattering screen is likely the ionized outer layers of molecular clouds \citep[][and references therein]{lc98b,lag+99}. This implies that scattering material is patchy with a more complicated spatial structure than a single thin screen.  The NE2001 model \citep{cl02} also implements the GC scattering region as an ellipsoid exponential. An analog to the complex GC scattering may be the time-variable scattering of the Crab pulsar due to the complex spatial structure in the Crab nebula, which imparts rapid changes in the scattering time scale \citep{kss10} and in extreme cases anomalous dispersion events \citep{bwv00}. Another possibility is the magnetar may be at a larger radial distance from Sgr A* but viewed through a filament, that may cause the higher DM but lower scattering, but the large measured RM argues against this interpretation. 
Continued high precision monitoring of the DM and RM of PSR J1745-2900 may show whether the source is moving near or within an extended screen boundary.

Still, an alternative scenario to be considered is a real paucity of pulsars in the GC, as suggested earlier by Johnston (1994)\nocite{joh94}. Newer results, including the discovery the discovery of PSR J1745--2900, contradict this possibility. Based on population and multi-wavelength studies, reviews of the physical conditions and considerations of the stellar population and indications of their formation history, the number of pulsars expected in the GC is in fact high \citep{lk04}. \citet{wcc+12}  predict as many as 100 canonical pulsars and a ten times larger population of millisecond pulsars in the interesting central parsec of the GC. Because radio magnetars are a rare class of pulsar (1 in $\sim$500 radio pulsars), this detection suggests an even larger population may be present. From this population PSR J1745--2900 is precisely the type of pulsar we expect to detect in a region of strong scattering through selection effects viz.\ high luminosity, long period, flat spectrum. If the scattering properties of the PSR J1745--2900 is indicative of the entire GC region, then previous surveys should have detected canonical pulsars. MSPs will not be detectable at low frequencies, even with the small scattering time measured here for the magnetar, but high frequency searches could have had the chance to discover some of them \citep{ekk+13}. Still, no discovery was made before the observations of the magnetar. This apparent contradiction between the low pulse broadening measured for PSR J1745-2900 and the lack of other pulsar detections suggests that the scattering environment in the GC requires a more complicated spatial model than a single thin scattering screen, and hyperstrong scattering may still dominate most lines-of-sight. 

As future searches of the GC with more sensitive telescopes (e.g. with the Square Kilometre Array, Atacama Large Millimeter Array) are made, and monitoring of the magnetar continues, the question of whether or how we can find pulsars closely orbiting Sgr A* will eventually be answered.

\section{Conclusions}
\label{conclusions}
We have observed the magnetar PSR J1745--2900 at radio frequencies ranging from 1.2 to 18.95 GHz. Like other radio-emitting magnetars, the average pulse profile of PSR J1745--2900 is comprised of bright, narrow pulses that jitter in time. Using both average pulse profiles and single pulses, we measured the scatter broadening timescale across an order of magnitude in observing frequency and found $\tauGHz = 1.3 \pm 0.2$ s and $\alpha = -3.8\pm 0.2$. The pulse broadening timescale is several orders of magnitude lower than predicted by models \citet{cl02} for a pulsar near Sgr A*. If there are truly as many as 100 canonical pulsars in the inner 1 pc around Sgr A*, then previous pulsar surveys should have detected many sources. Our models for scattering in the GC require updating after the discovery of this magnetar.

\acknowledgements
L.~G.~Spitler and R.~Karuppusamy gratefully acknowledges the financial support by the European
Research Council for the ERC Starting Grant BEACON under contract no.
279702. M.~Kramer, K.~J.~Lee, and C.~G.~Bassa gratefully acknowledge support from ERC Advanced 
Grant ``LEAP'',Grant Agreement Number 227947 (PI Michael 
Kramer). The authors would like to thank Bill Coles for useful discussions.

%\bibliography{gc_magnetar}

\end{document}